\documentclass[fleqn,10pt]{wlscirep}
\usepackage[utf8]{inputenc}
\usepackage[T1]{fontenc}

\usepackage{subcaption}

\title{Enabling Additive Manufacturing Part Inspection of Digital Twins via Collaborative Virtual Reality}

\author[1,*]{Vuthea Chheang}
\author[1]{Saurabh Narain} %narain1@llnl.gov
\author[1]{Garrett Hooten} %hooten1@llnl.gov
\author[1]{Robert Cerda} %cerda3@llnl.gov
\author[1]{Brian Au} %au7@llnl.gov
\author[1]{Brian Weston} %weston8@llnl.gov
\author[1]{Brian Giera} %giera1@llnl.gov
\author[1]{Peer-Timo Bremer} %bremer5@llnl.gov
\author[1]{Haichao Miao} %miao1@llnl.gov
\affil[1]{Lawrence Livermore National Laboratory, Livermore, CA, 94550, United States}

\affil[*]{chheang1@llnl.gov}


\begin{abstract}

Digital twins (DTs) are an emerging capability in additive manufacturing (AM), set to revolutionize design optimization, inspection, in situ monitoring, and root cause analysis. AM DTs typically incorporate multimodal data streams, ranging from machine toolpaths and in-process imaging to X-ray CT scans and performance metrics. Despite the evolution of DT platforms, challenges remain in effectively inspecting them for actionable insights, either individually or in a multidisciplinary team setting.
Quality assurance, manufacturing departments, pilot labs, and plant operations must collaborate closely to reliably produce parts at scale. This is particularly crucial in AM where complex structures require a collaborative and multidisciplinary approach. 
Additionally, the large-scale data originating from different modalities and their inherent 3D nature pose significant hurdles for traditional 2D desktop-based inspection methods. To address these challenges and increase the value proposition of DTs, we introduce a novel virtual reality (VR) framework to facilitate collaborative and real-time inspection of DTs in AM. This framework includes advanced features for intuitive alignment and visualization of multimodal data, visual occlusion management, streaming large-scale volumetric data, and collaborative tools, substantially improving the inspection of AM components and processes to fully exploit the potential of DTs in AM.

\end{abstract}

\begin{document}

\flushbottom
\maketitle

%
%  Click the title above to edit the author's information and abstract
%
\thispagestyle{empty}

\begin{figure}[t]
\centering
\includegraphics[width=\linewidth]{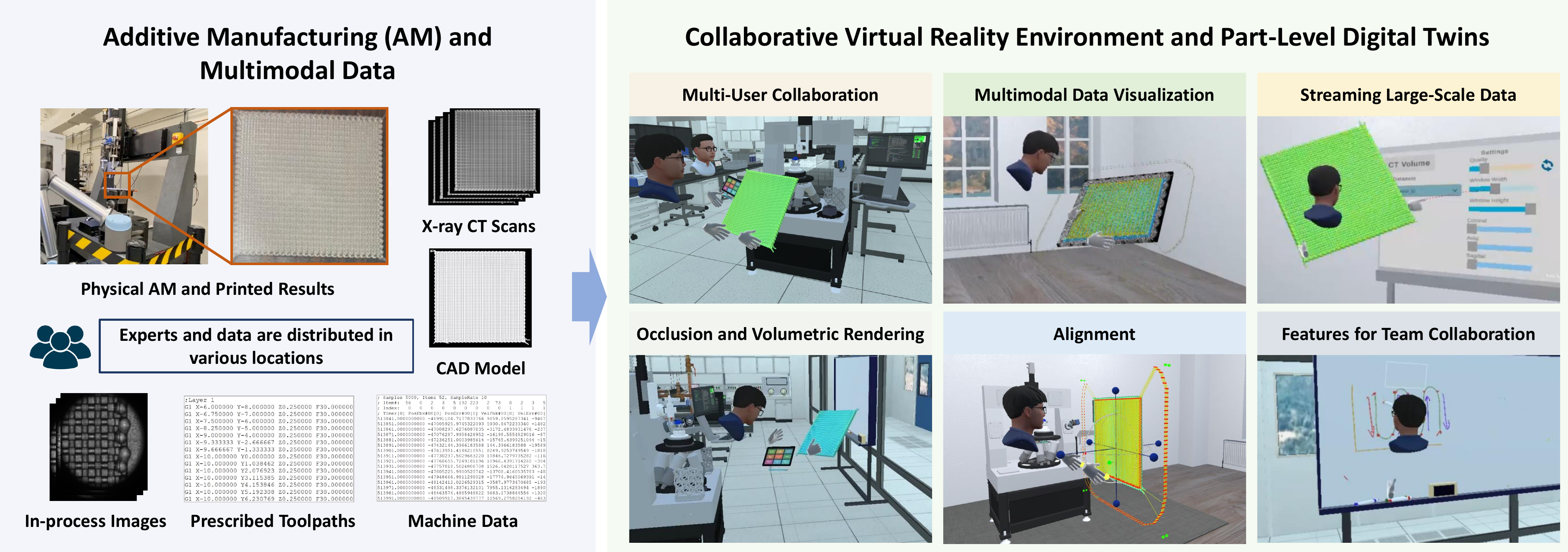}
\caption{Overview of the proposed framework aimed to enhance AM part inspection of DTs via collaborative VR. It supports multimodal data alignment and visualization, streaming large-scale and multi-resolution volumetric data, visual occlusion management, as well as team collaboration features. Furthermore, it allows multiple users either co-located or remote to collaborate in a shared virtual environment.     
}
\label{fig:overview}
\end{figure}

\section*{Introduction}

Additive manufacturing (AM) has been revolutionizing the production of complex parts across various industries. 
AM involves creating an object in a layerwise~\cite{stawski20233d} or volumetric fashion~\cite{kelly2019vol}, using either polymeric, ceramic, metallic, or multi-material printing. 
A critical challenge common to all AM techniques is to ensure that the final part meets the design criteria. Defects in AM, such as excessive or insufficient material deposition, internal voids, or porosity defects are often concealed inside the part and not accessible to traditional inspection methods~\cite{brennan2021defects, maconachie2019slm}.  
Detecting such defects requires expensive and time-consuming computed tomography (CT) scans which are subsequently challenging to process due to their size, complexity, and inherent spatial nature. Nevertheless, inspection is an unavoidable step in the overall manufacturing process.

Given the ubiquitous challenges associated with process monitoring, part inspection, and part performance verification, industrial and commercial sectors are building digital twins (DTs) of AM platforms to mitigate these issues~\cite{tao2018digital, attaran2023digital, gunasegaram2021case}. While there exist various definitions of what constitutes a DT in AM, the most common notions include a variety of rich multimodal data streams covering the entire production cycle from design to printing, to inspection and lifecycle management. In this context, AM data is collected from various sources, such as design files, sensing suites, images, prescribed toolpaths, machine kinematics and health monitoring, CT scans, and other performance measurements.
However, few approaches exist that can automatically and jointly process all different modalities and often the first challenge users face is how to productively explore the DT data in a coherent fashion~\cite{lu2020digital, liu2021review}. Here, we contend that carefully designed visualizations and intuitive interactions coupled to a shared virtual environment can make DTs significantly more effective. 
Nevertheless, designing such visualizations for the data-intensive and multimodal streams %we expect and doing so 
in an intuitive and scalable manner remains a challenge~\cite{gunasegaram2021towards, liu2022digital_dt}.
For example, CT scans, while indispensable for ensuring the structural integrity and accuracy of these complex parts, produce large-scale and complex volumes. 
The three-dimensional and dense nature of these datasets makes them not only difficult to inspect but also challenging to interpret, especially when considering the boutique or low-volume production typical of AM applications.  
In addition, experts from design, production, and related teams need to be involved to ensure its accuracy, reliability, and usability~\cite{scime2022scalable, phua2022digital, havard2019digital}.

Traditional approaches and desktop tools provide only limited visualization and interaction opportunities for effectively handling the spatial nature of AM components and their DT representations~\cite{pantelidakis2022digital, yan2023digital}. 
Compared to desktop-based systems, virtual reality (VR) provides benefits to deal with several challenges in AM, including optimizing process, maintenance, part quality, and inspection~\cite{akpan2024role, kobara2023bibliometric}. 
The spatial nature of AM parts and processes requires a new set of capabilities by leveraging the advantages of VR to facilitate a more intuitive and immersive inspection process and meet the specific demands of AM, such as multimodal data handling and alignment, occlusion management for inspecting internal defects, streaming large-scale volumetric data, and real-time synchronization for distributed expert collaboration~\cite{wang2023ontology, pirker2022immersive, litvinova2018collaborating}.

In response to these challenges, we introduce a novel collaborative VR framework designed to enhance the inspection and analysis of AM parts through multimodal data (see~\autoref{fig:overview}). Our framework not only fosters collaborative interactions among geographically distributed users but also incorporates several innovative approaches aimed at improving the understanding and inspection of complex AM parts and processes. These features include intuitive alignment capabilities within the VR environment, visualization strategies for effective visual occlusion management, and VR-based volume rendering of the volumetric data. Moreover, our framework provides the ability to seamlessly stream large-scale volumetric data addressing the critical issue of handling the massive datasets generated by CT scans, further augmented by features for team collaboration, e.g., annotation functionalities, that enhance the collaborative inspection process. 
%The development of this VR-based environment signifies a paradigm shift in the inspection and analysis of AM components. By removing the barriers to effective collaboration across geographically distributed experts, e.g., designers, manufacturers, and customers, and enabling a data-informed communication model using a virtual environment, we not only improve the efficiency of distributed manufacturing processes but also set a new standard for the development and integration of a DTs ecosystem. 

%In this work, we DO GREAT STUFF DESCRIBED IN 2 SENTENCES.
%In our work, we address the critical needs of emerging DTs by introducing an innovative collaborative VR framework specifically designed for the demands of inspection and analysis in AM. 
Our framework is designed to align seamlessly with the spatially complex, multimodal data integral to AM DTs, fostering novel team collaboration. 
It represents a paradigm shift in how users interact with, analyze, and derive actionable insights from complex manufacturing data, setting a new benchmark in this emergent field and pushing the boundaries of current DT inspection methods.

%By addressing the critical challenges of data complexity, distributed expertise, and the need for scalable inspection solutions, our framework provides a new benchmark and opens research directions for efficiency, collaboration, and innovation in AM processes and manufacturing.

\section*{Materials and Methods}

AM DTs are described by multimodal data streams collected from the processes, including prescribed toolpaths used for instructing the printer, machine toolpaths describing the actual printing locations, X-ray CT used for investigating internal material densities, and in-process images for layerwise verification. 
To provide a comprehensive inspection tool for those modalities, however, there are multiple steps to achieve this goal. We first need to align those modalities and provide multimodal data visualization. Since X-ray CT scans are crucial for inspection, occlusion management plays a vital role in conveying depth information of volumetric rendering for inner-structure inspection. Moreover, X-ray CT scans are often large and challenging to manage and share between users. Hence, a solution is needed to store and provide flexible data management.
Team collaboration is not only sharing a 2D screen, but the users need to step into the shared immersive environment to explore and inspect those DT representations. 

In the following sections, we describe collaborative VR-based approaches designed and developed to enhance the inspection and analysis of AM parts. 
The proposed framework allows individual or distributed users to perform data exploration and inspection in the shared virtual environment via real-time synchronization. Moreover, the framework incorporates several new approaches:  \textit{intuitive alignment}, \textit{multimodal data visualization}, \textit{comprehensive occlusion management}, \textit{streaming large-scale volumetric data}, and \textit{features for synchronous team collaboration}.

\begin{figure*}[t] %0.323
  \centering
    \begin{subfigure}{0.242\textwidth}
        \includegraphics[width=\columnwidth]{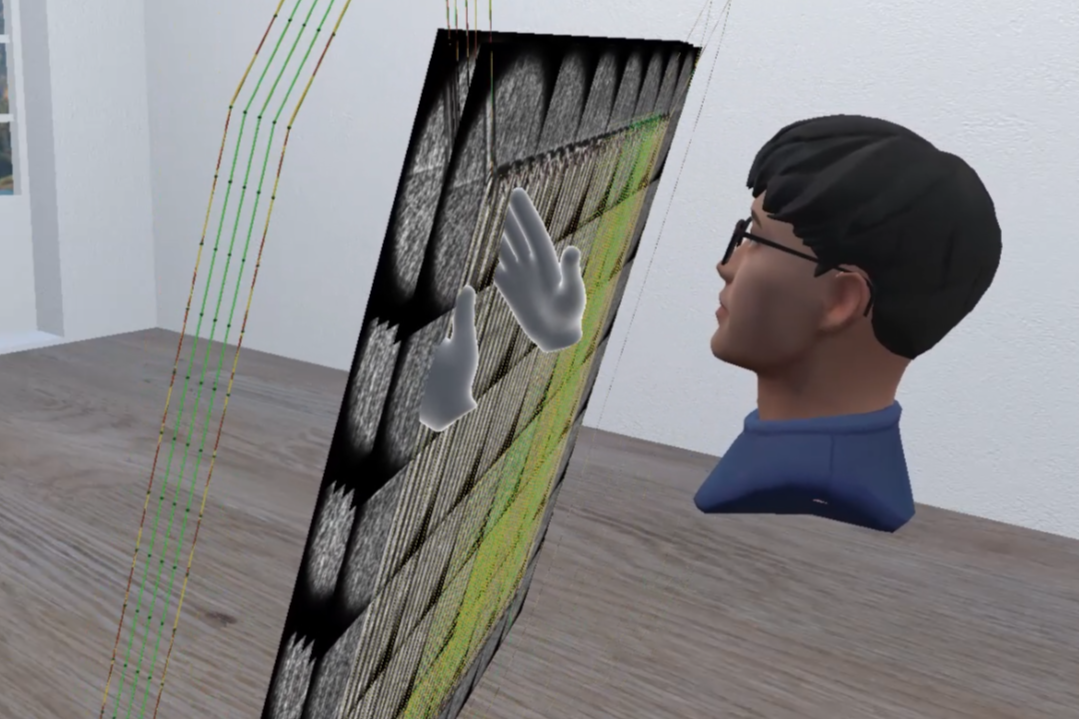}
        \caption{}
        \label{fig:multimodal1}
    \end{subfigure}
    \begin{subfigure}{0.242\textwidth}
        \includegraphics[width=\columnwidth]{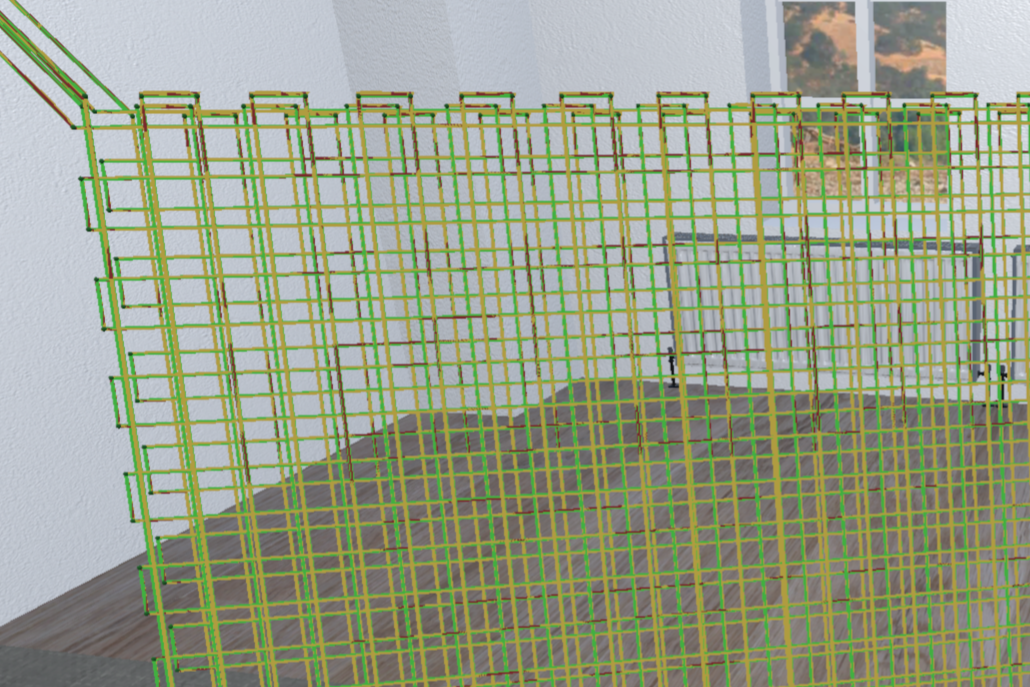}
        \caption{}
        \label{fig:multimodal2}
    \end{subfigure}
    \begin{subfigure}{0.242\textwidth}
        \includegraphics[width=\columnwidth]{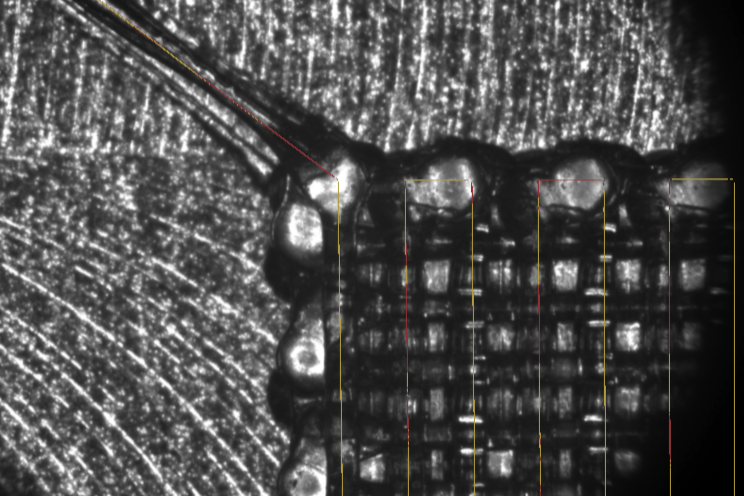}
        \caption{}
        \label{fig:multimodal3}
    \end{subfigure}
    \begin{subfigure}{0.242\textwidth}
        \includegraphics[width=\columnwidth]{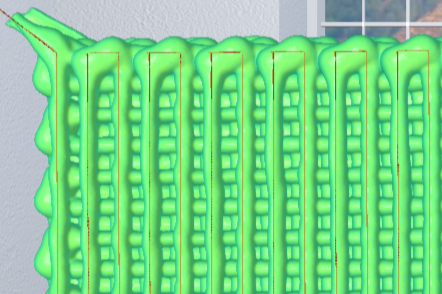}
        \caption{}
        \label{fig:multimodal4}
    \end{subfigure}
  \caption{Multimodal data visualization to support the process of data inspection and analysis: (a) users can load and explore data from multiple data streams in the virtual environment, (b) data representations based on printed layer, including prescribed toolpath (green), machine toolpath (yellow), and error (red), (c) users can enable in-process imaging to further inspect the data, and (d) volumetric data visualization from X-ray CT scans with machine toolpaths.}
  \label{fig:multimodaldata}
\end{figure*}

\subsection*{Intuitive Alignment}

Comparing different modalities of AM DTs is crucial for inspection, e.g., comparing between prescribed and actual toolpaths or X-ray CT scans after the printing to the intended design. Valuable insights can be gained from these comparisons to highlight and inspect deviations between modalities and potential defects. 
The fundamental challenge prevalent in many DT applications is that each modality has its own format, units, and coordinate systems. 
Without proper alignment, those modalities cannot be compared and integrated effectively.   
% For instance, isosurface and X-ray CT data may undergo registration during setup and processing. 
For instance, additional adjustments are required to align the prescribed toolpaths, machine toolpaths, in-process images, and the X-ray CT volume to ensure accurate visualization and interpretation. 
In this case, inspecting how a location in one modality, i.e., in-process images, overlays in another modality, i.e., X-ray CT volume, would not be possible without properly aligning the data. 
We believe that the VR-based alignment could be a crucial tool to optimize these modalities' visualization and support the multimodal analysis.

While mid-air interaction is a common approach for interacting and manipulating objects in VR, such interactions could also suffer from %imprecise controller tracking and 
hand instability, e.g., unnoticeable tremor, which introduces inaccuracies when trying to align at a high precision. 
To achieve this goal, we developed a VR-based alignment tool with an adaptation and enhancement of a precise alignment technique proposed by Rodrigues et al.~\cite{rodrigues2023amp}. Instead of a direct linear mapping between controller and object movement, this approach utilizes a non-linear mapping that adapts to the acceleration of the controller to determine the object movement, e.g., the object moves more where the controller is moving faster and vice versa. With slow controller movements mapping to small object movement, we utilize this technique to achieve high-precision alignment of multimodal data. 
Since the printing process often presents both local and global defects that go beyond the rigid body alignment, we further integrated a free-form deformation technique~\cite{sederberg1986free} to provide a flexible and intuitive alignment for non-linear deformation.
It first defines the control points based on the bounding box or grid structure of the model. Each control point has its position and acts as a handle for deforming the object. The user can adjust the deformation by interacting with the control points. Once the control point is moved, the interpolation method is computed for affected nearby vertices to smoothly deform the object. These transformations can be saved for revisiting the alignment or exporting to other tools for further registration.

\subsection*{Multimodal Data Visualization}

To fully understand and analyze the complex behavior of AM DTs, relying solely on a single type of data visualization is often insufficient. 
While there are possibilities to gather new forms of data from physical assets, the ability to process and visualize them into valuable insights and situational awareness, such as corresponding locations in different modalities, is challenging as one modality would visually occlude another when they are superimposed. 
%Multimodal data visualization in the immersive environment allows users to understand the complex relationship between various data modalities.  
%It is extremely useful for the inspection and analysis process. 
The main benefit of multimodal data visualization for AM inspection and analysis is to provide a comprehensive comparison across different data streams, e.g., before vs after, pre-cure vs post-cure, or nominal vs actual printing part.
One possible solution is to superimpose the visualization of one modality over another, e.g., computer-aided design (CAD) model and X-ray CT scans~\cite{klacansky2022virtual}. However, this introduces the issues of occlusion and robustness between modalities. 

In this work, we incorporate several modalities, ranging from toolpaths and in-process images to X-ray CT scans. 
For instance, overlaying additional modalities, i.e., machine toolpaths on top of the CT scans and in-process images can help identify areas of defects and provide a more comprehensive understanding of the printing process and behavior (see \autoref{fig:multimodaldata}). 
Current supported data streams include design models, prescribed toolpaths (\textit{.pgm}), machine data (\textit{.hdf5)}, including machine toolpaths and errors, in-process images, X-ray CT scans, and 3D reconstructed models.
The proposed framework can be used to load those data types and visualize them in the immersive virtual environment. 
Hence, users can intuitively explore, analyze, and gain a comprehensive understanding of the process, which can make informed decisions to optimize the process and quality. % and reduce defects. 
We also implemented options to support multimodal exploration and analysis. The users can inspect it layer by layer, enable/disable one of the data streams, and compare it with different modalities. 
As shown in \autoref{fig:multimodaldata}, prescribed toolpaths, which are used to instruct the 3D printer (\textit{G-Code commands}) are visualized as green lines with corresponding positions (green dots). 
The actual paths generated by the 3D printer are visualized with yellow lines, and errors between prescribed and machine toolpaths are displayed in red. 
The users also have options to enable and overlay in-process images as well as volumetric rendering of the X-ray CT scans for further inspection and analysis.  

\begin{figure*}[t] %0.323
  \centering
    \begin{subfigure}{0.194\textwidth}
        \includegraphics[width=\columnwidth]{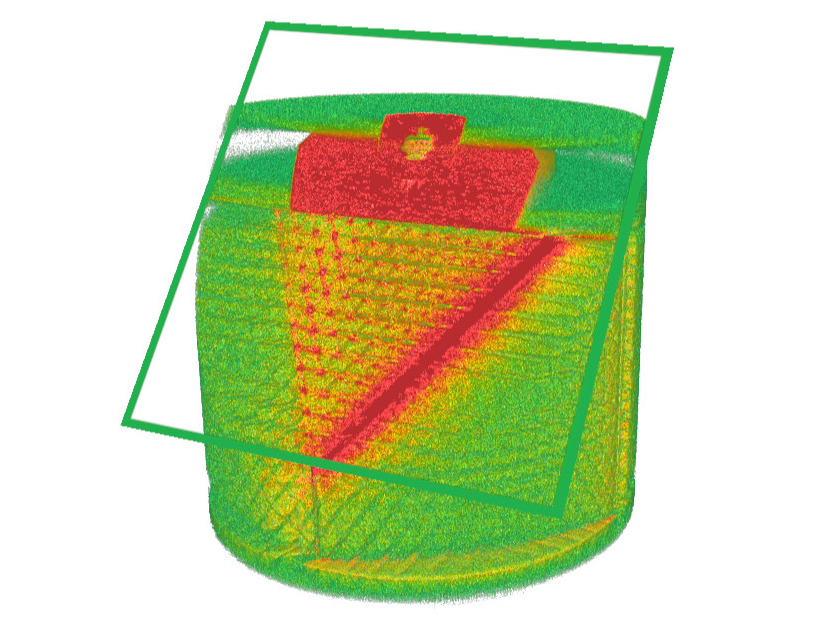}
        \caption{}
        \label{fig:occulsion0}
    \end{subfigure}
    % \begin{subfigure}{0.194\textwidth}
    %     \includegraphics[width=\columnwidth]{Figures/occlusion_1.png}
    %     \caption{}
    %     \label{fig:occulsion1}
    % \end{subfigure}
    % \begin{subfigure}{0.194\textwidth}
    %     \includegraphics[width=\columnwidth]{Figures/occlusion_3.png}
    %     \caption{}
    %     \label{fig:occulsion3}
    % \end{subfigure}
    % \begin{subfigure}{0.194\textwidth}
    %     \includegraphics[width=\columnwidth]{Figures/occlusion_5.png}
    %     \caption{}
    %     \label{fig:occulsion5}
    % \end{subfigure}
    \begin{subfigure}{0.194\textwidth}
        \includegraphics[width=\columnwidth]{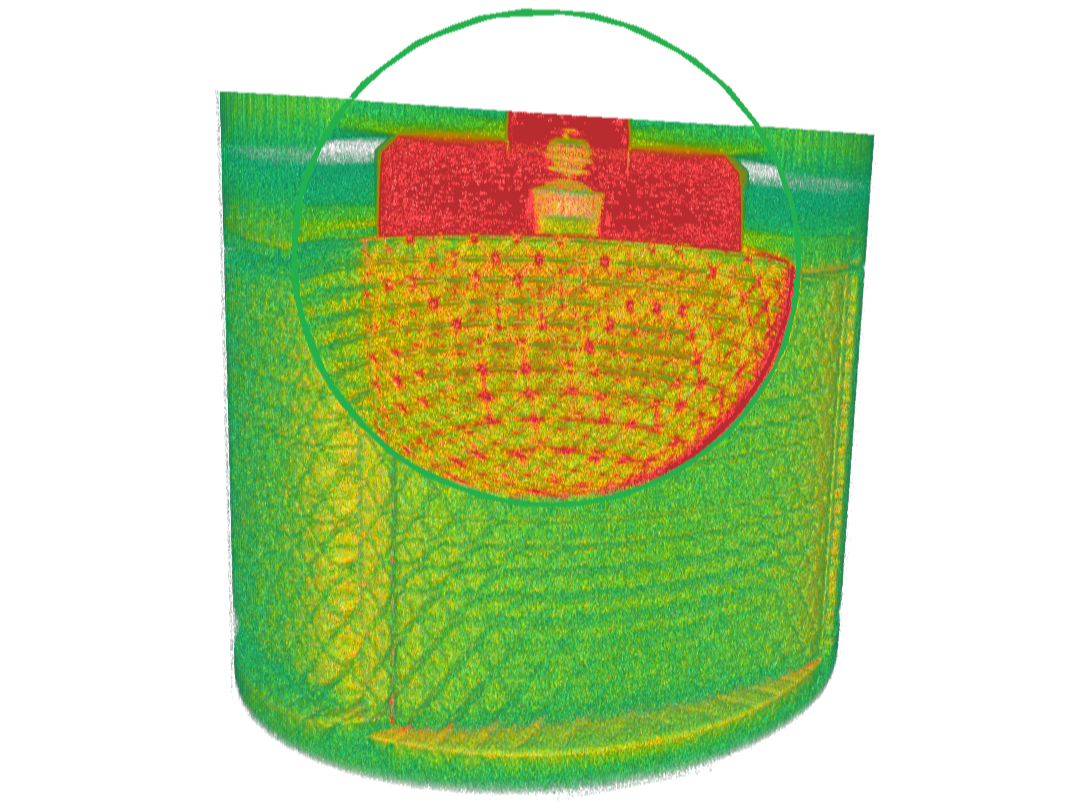}
        \caption{}
        \label{fig:occulsion7}
    \end{subfigure}
    \begin{subfigure}{0.194\textwidth}
        \includegraphics[width=\columnwidth]{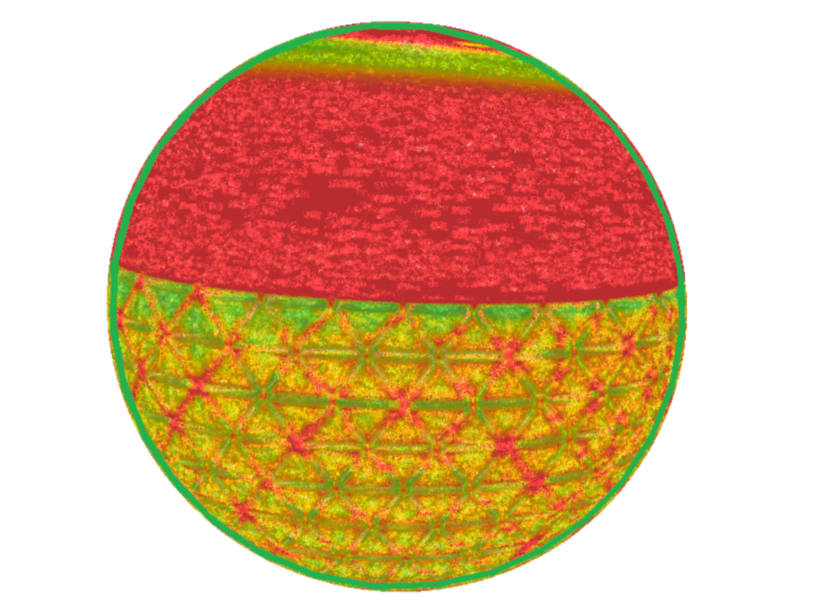}
        \caption{}
        \label{fig:occulsion9}        
    \end{subfigure}
    \begin{subfigure}{0.194\textwidth}
        \includegraphics[width=\columnwidth]{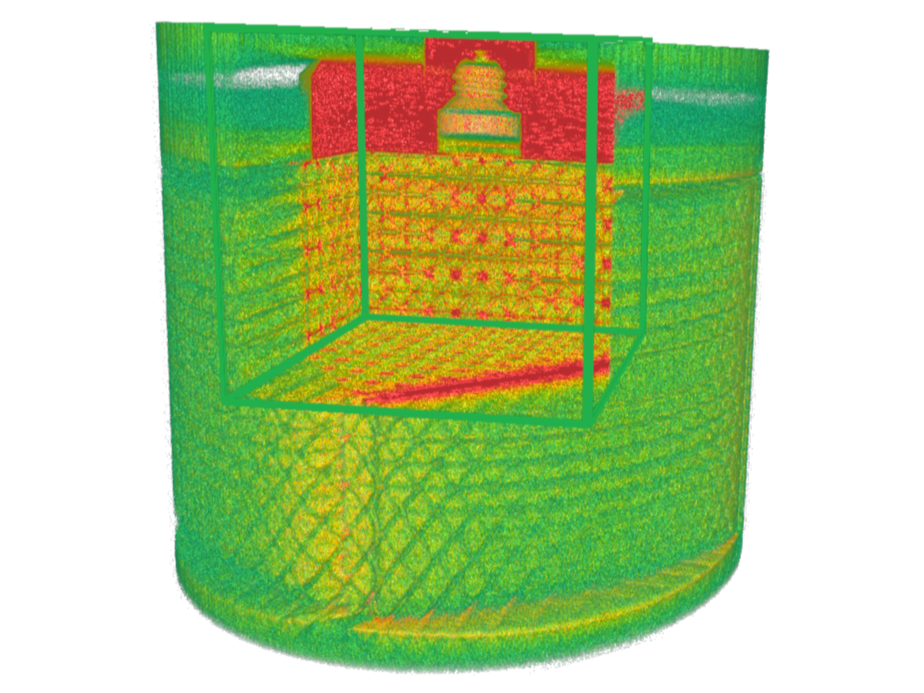}
        \caption{}
        \label{fig:occulsion8}
    \end{subfigure}    
    % \begin{subfigure}{0.194\textwidth}
    %     \includegraphics[width=\columnwidth]{Figures/occlusion_2.png}
    %     \label{fig:occulsion2}
    % \end{subfigure}
    % \begin{subfigure}{0.194\textwidth}
    %     \includegraphics[width=\columnwidth]{Figures/occlusion_4.png}
    %     \label{fig:occulsion4}
    % \end{subfigure}
    % \begin{subfigure}{0.194\textwidth}
    %     \includegraphics[width=\columnwidth]{Figures/occlusion_6.png}
    %     \label{fig:occulsion6}
    % \end{subfigure}    
    \begin{subfigure}{0.194\textwidth}
        \includegraphics[width=\columnwidth]{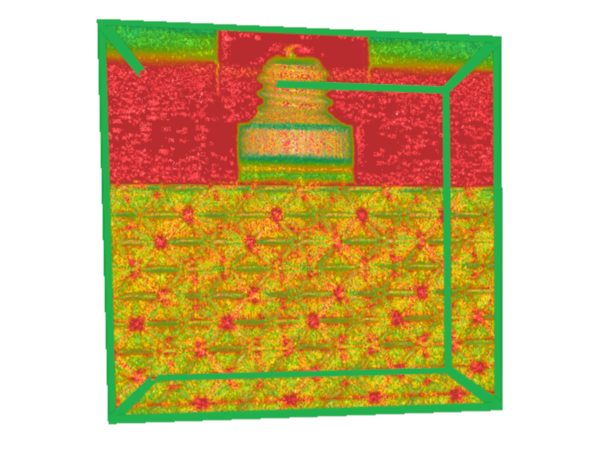}
        \caption{}
        \label{fig:occulsion10}        
    \end{subfigure}
   \caption{Occlusion management with volumetric rendering in immersive VR. We implemented the colorized volume with the transfer functions and color maps.
   With the cutting objects, it allows users to explore the inner structures of the volumetric data smoothly, e.g., using a cross-section plane (a), sphere (b and c), or box (d and e) cutout with inclusive and exclusive modes, respectively.}
  \label{fig:occulsion}
\end{figure*}

\subsection*{Occlusion Management and Volumetric Rendering}

Volume rendering is used to visualize final part volumetric data obtained from X-ray CT scans. Since our AM lattice strucutres are dense, it is challenging to distinguish and identify the inner-structure details of volumetric data without occlusion management.  
It is also crucial to enhance the depth perception and shape of the rendered object exposed to its surrounding environment. 
%In direct volume rendering, ambient occlusion can be computed in each voxel by calculating how much the neighboring voxels obscure with it~\cite{titov2024contextual, kroes2018smooth}. 
To reduce occlusion in volumes, a common approach is to apply a transfer function allowing the user to specify structures to be visualized and manipulate the opacity and color of the voxels belonging to the structures~\cite{tian2018occlusion}. 
For instance, the transfer function can be used to control the visualization of the structures that occlude the regions of interest.
However, adjusting only the transfer function does not allow for global adjustments of visibility.  

We enhanced \textit{UnityVolumeRendering} %\footnote{https://github.com/mlavik1/UnityVolumeRendering} 
to compute and visualize volumetric data in \textit{Unity}. 
The transfer function and color mapping were adapted and integrated into the proposed framework to improve the occlusion and lighting management in VR. 
%Three datasets from X-ray CT scans of AM parts were tested. %A comparison between the volumetric rendering with and without occlusion management is shown in \autoref{fig:occulsion}. 
To have a global way to adjust visibility and inspect the inner structure of volumetric data while maintaining local adjustments, we integrated cutting geometries, such as cross-section plane, box, and sphere cutouts similar to the technique proposed by Titov et al.~\cite{titov2024contextual}.
The cutout function supplements with inclusive and exclusive modes, which can either make the inside visible or invisible (see \autoref{fig:occulsion}).
This way allows for a flexible adjustment of visibility, thus, the inner structures inside the volumetric data can be explored and inspected.
The users in VR can interactively move the cutting object around and choose the cutout mode to crop the region of interest, e.g., to further inspect and discuss the data with their collaborators accordingly.

\subsection*{Streaming Large-Scale Volumetric Data}

%The X-ray CT scans of AM parts are often large, which could result in high-resolution volumes. 
%However, storing datasets locally, especially for collaborative VR users, could pose concerns regarding data security and privacy. 
%However, storing and sharing large-scale data among collaborators is challenging.
%Providing high frame rates while dealing with resource-intensive computation is also challenging to avoid discomfort for VR users. 

X-ray CT scans are crucial for the AM inspection process to analyze defects and ensure the overall quality and reliability of the printed parts. They provide a non-invasive and non-destructive way to characterize and inspect internal material densities of the AM parts, particularly complex internal structures that are not accessible by other inspection methods.
However, X-ray CT scans are often large, which cannot be easily shared with collaborators and also impose computation and rendering challenges for VR.

One of the key features of our proposed framework is the ability to utilize large-scale data management and stream multi-resolution volumetric data remotely during runtime. 
%We utilized \textit{OpenViSUS}\footnote{https://github.com/sci-visus/OpenVisus}, which is an open-source library used for efficient querying and data management of large-scale datasets. 
We utilized the data management approach using \textit{OpenViSUS} for efficient querying and managing large-scale data~\cite{zhou2023orchestration, pascucci2012visus}. 
The datasets are stored on a server and can be streamed with our developed \textit{WebAPI} to provide flexible data streaming, e.g., filtering, cropping, and multi-resolution data query. 
This way, we can instantly get a subsampled version or a high-resolution subspace of the data, so that interactive inspection is guaranteed and not inhibited by the large-scale nature.
Furthermore, we incorporated the development with multithreading to support the massive computations in the background. Thus, the main thread could be used to maintain the computational logic for VR rendering.

\begin{figure}[t]
\centering
\includegraphics[width=\linewidth]{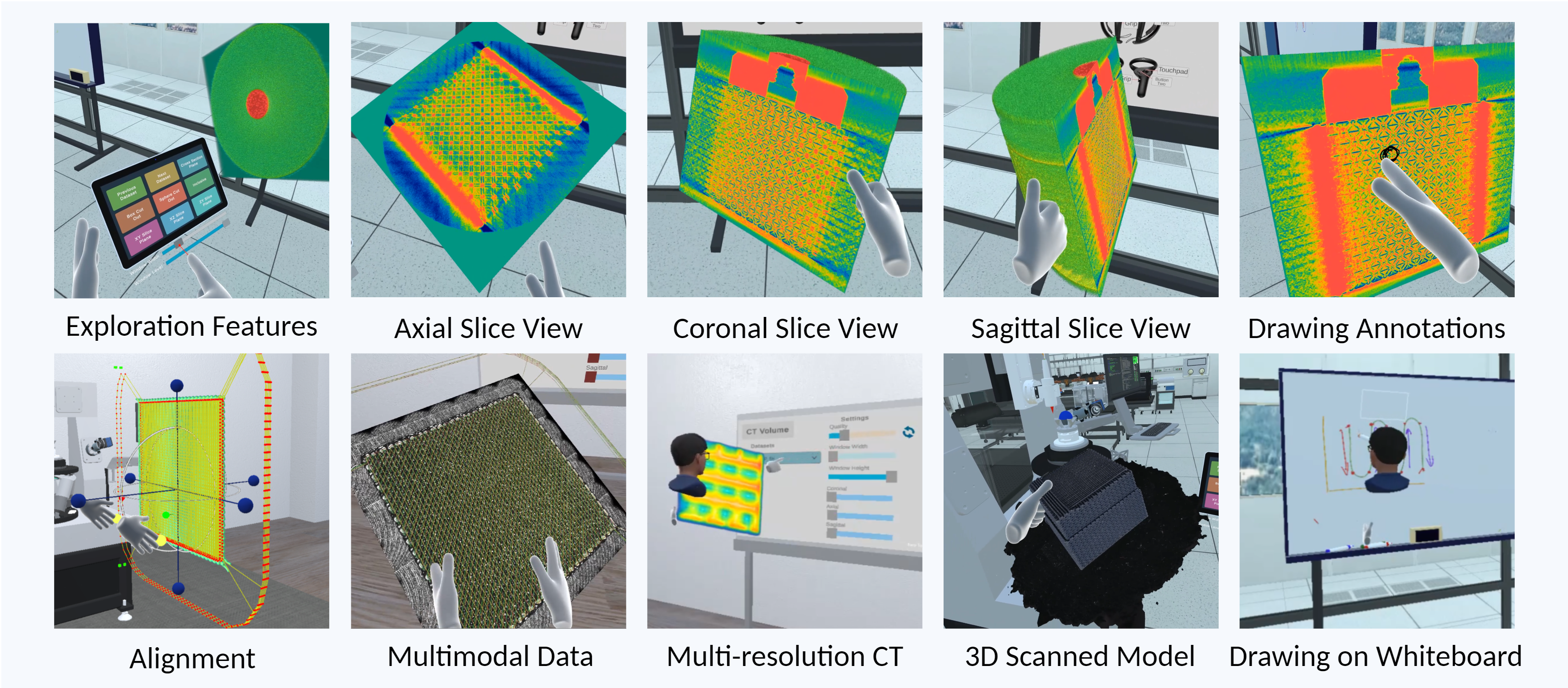}
\caption{Various features for data exploration and inspection as well as related features to improve team-based collaboration: those features include adjusting CT window width/level, using cross-section plane and axis slice view (axial, coronal, and sagittal) to explore and inspect data, drawing annotations, aligning and visualizing multimodal data, streaming multi-resolution CT volume, loading and interacting 3D scanned models, and drawing on the virtual whiteboard.
}
\label{fig:features}
\end{figure}

%\subsection*{Features for Team Collaboration}

\subsection*{Multi-User and Collaborative Inspection} %Collaborative Data Exploration and Inspection

%Traditional approaches often require physical presence, which can be time-consuming and costly, back-and-forth travel, and might be restricted during the global pandemic. Moreover, using desktop-based systems may not provide the level of detail necessary for effective inspection of additively manufactured parts. 

Conventionally, designers, production teams, domain experts, and sometimes customers directly interact with the physical objects after the AM part has been produced, often requiring the shipping of physical components or experts traveling to different locations. 
One notable characteristic of the parts produced through AM is the intricate internal geometric structures. 
Inspecting these structures requires volumetric imaging, i.e., CT scans that lead to the well-known challenges of visualizing it through 2D interfaces, and domain knowledge by multiple experts~\cite{klacansky2022virtual, pirker2022immersive}. 
% Compared to desktop-based systems, VR offers revolutionary possibilities for data exploration and inspection by providing more immersive and intuitive interactions, improving spatial awareness of the part, and increasing the efficiency of the inspection. 
Collaborative VR offers extra benefits by enabling real-time communication, jointly shared inspection and analysis, and facilitating team-based discussion.

The proposed framework provides exploration and inspection features via real-time synchronization. \autoref{fig:features} shows features developed to support the team in exploring and inspecting AM parts.   
A virtual tablet was designed to provide user interface interactions. The users can change different datasets, adjust volume intensity, i.e., CT window width and level, and use data exploration features, such as cross-section plane and slicing views from different axes, including axial, coronal, and sagittal views.
Drawing annotations in collaborative VR can be essential for team discussion. It can be used to enhance the collaboration between team members more effectively.
The users can draw annotations to specify the region of interest on the dataset and initiate the discussion accordingly. 
The drawing was implemented by using a VR ray casting technique mapped with the index fingertip of the virtual hand.
Moreover, we developed a virtual whiteboard to support the team discussion. The users can use a virtual marker to draw their illustrations on the whiteboard in a similar way that they perform in the physical world. 

\begin{figure*}[t]
  \centering
    \begin{subfigure}{0.485\textwidth}
        \includegraphics[width=\columnwidth]{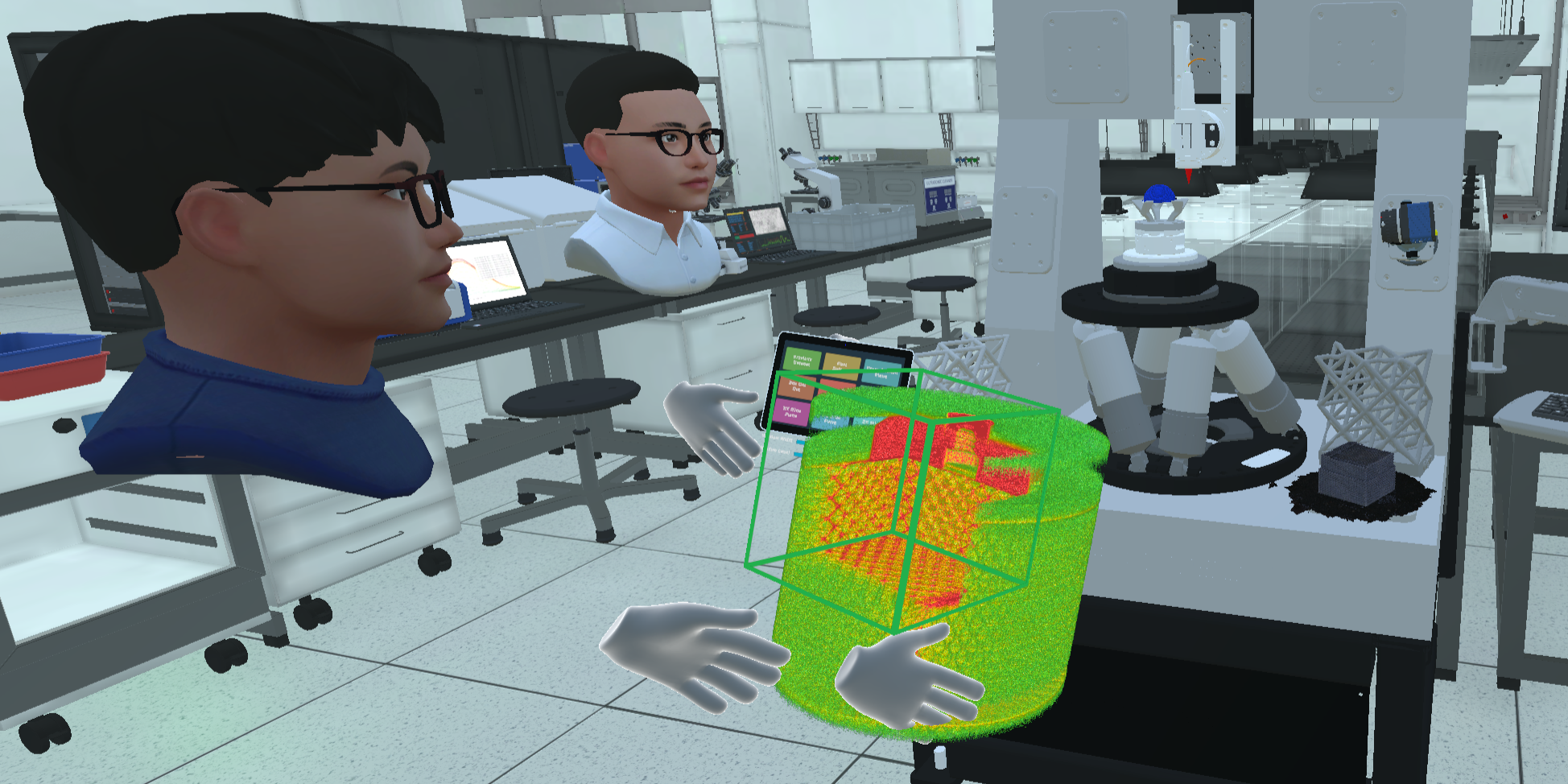}
        \caption{}
        \label{fig:inspection1}
    \end{subfigure}
    \begin{subfigure}{0.485\textwidth}
        \includegraphics[width=\columnwidth]{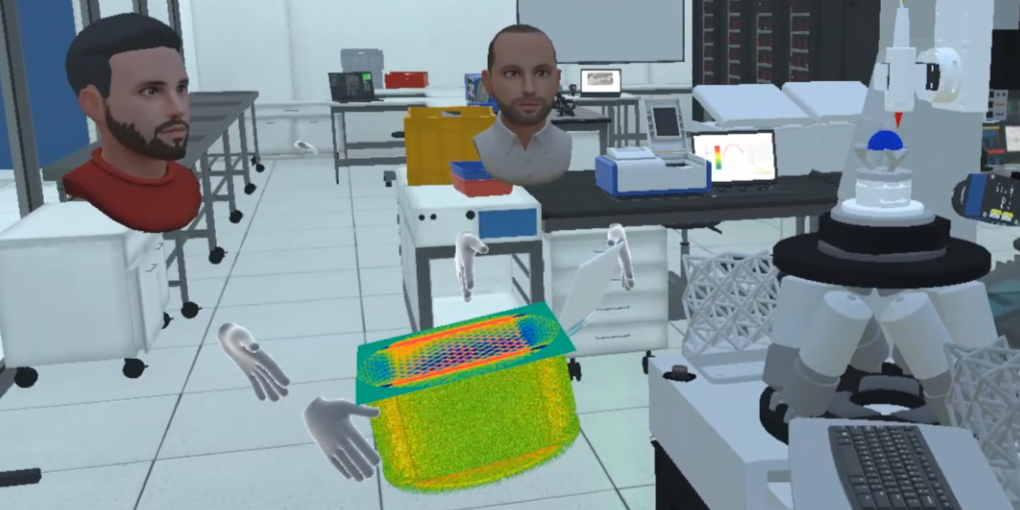}
        \caption{}
        \label{fig:inspection2}
  \end{subfigure}
  \caption{Multi-user collaboration for AM data exploration and inspection: (a) collaborative users virtually join and explore AM data, i.e., X-ray CT volume, in a shared virtual environment, and (b) users can inspect further details of the data using various inspection features, i.e., a cross-section plane. The interactions between users are synchronized in real time.}
  \label{fig:collaborativeinspection}
\end{figure*}

To provide an engaged environment for team communication, we %utilized a \textit{Ready Player Me} library to 
design personalized user avatar representations, including an avatar head generated from a headshot, animated VR hands, facial blendshapes, and the voice icon that appears when the user engages in voice communication.
\autoref{fig:collaborativeinspection} shows collaborative users explore and inspect AM data in the virtual environment.
While the environment is designed and developed for VR users using head-mounted displays (HMD), non-VR users can also join in the environment as a spectator mode using conventional inputs such as mouse and keyboard. 
This feature would be essential to allow other users, e.g., trainees, to observe and quickly join the discussion in the virtual environment.

We developed the collaborative VR environment using a client-server approach. 
For instance, a client sends an update through the server, and the server multicasts the update to other clients for real-time synchronization.
This approach could provide a stable and secure solution for network communication and avoid common connection issues compared to the peer-to-peer approach, e.g., scalability and over-distance connection. 
Furthermore, it is beneficial to allow clients who join late or are required to reconnect in the environment to load previous synchronization updates through the server. 
A \textit{Unity} game engine (Unity Software Inc., CA, USA) version \textit{2019.4.20f1} was used as a development environment. 
We used a \textit{Photon Unity Networking} (Exit Games GmbH, Germany) to provide load-balancing services and shared network sessions between the users. \textit{Photon Voice 2} was also utilized for voice communication. 
Since the client's computer specification and network conditions could affect the connections, we implemented the mechanism for data synchronization suggested by Singhal and Zyda \cite{singhal1999networked} to optimize the network latency. 
A \textit{remote procedure call} (RPC) approach was used to send requests and distribute data.
Sending large amounts of data in one chunk is avoided. 
To prevent the server's bottleneck, clients are also responsible for individual graphical rendering.  
Moreover, the clients are responsible for storing object states locally, while only sending updates of their periodic data, e.g., the object's positions, rotations, and scales during the interaction and active state.

\paragraph{Apparatus} Two VR-ready computers were used during the testing and evaluation. They were \textit{RAZER BLADE 16} laptops equipped with 13th Gen Intel(R)Core(TM)i9-13950HX (32 CPUs) processor, NVIDIA GeForce RTX 4090 (16GB VRAM) graphics card, and 32GB\,RAM. One user connected with a VIVE XR Elite VR headset, which has 1920\,$\times$\,1920 pixels per eye (3840 x 1920 pixels combined), 110-degree field of view (FOV), and 90\,Hz refresh rate. Another user used a Meta Quest 3 headset that has 2064\,$\times$\,2208 pixels per eye, 110-degree FOV, and up to 120\,Hz refresh rate. Both VR headsets were connected as PC VR. 
The speed of internet connection was approximately 247.58\,Mbps (download) and 248.41\,Mbps (upload).

\section*{Evaluation and Expert Feedback} %Results and Discussion

This work was reviewed and approved by Lawrence Livermore National Laboratory's review committee under \textit{LLNL-JRNL-861091}.
All methods were carried out in accordance with relevant guidelines and regulations.
To demonstrate the usability of our framework, we performed evaluations with domain experts.
Two exploratory studies were conducted: the first study aimed to assess the \emph{applicability and potential benefits} of the proposed framework, and the second study focused thoroughly on a \emph{case study} of inspection and analysis with real data collected from AM processes. 
The first study was conducted with six domain experts (three were computer scientists and the other three were engineering and material scientists). They all had experience working with AM. 
One reported having nine years of working experience, three reported between three to five years, and the other two rated as having between one to two years of working experience. 
The second study was conducted with two AM experts (two and six years of working experience). 
Exploratory and semi-structured interviews were conducted to assess the proposed framework and collect their qualitative feedback.
All the participants were informed about the objective and procedure of the study, and their verbal informed consent was collected.

\paragraph{Applicability and Potential Benefits}
All experts expressed their positive feedback. They confirmed the potential benefits and its applicability.
The framework can be a powerful tool to enhance the inspection process of AM DTs. 
The experts stated that using collaborative VR is beneficial to visualize and guide their collaborators, e.g., between the design and production teams, through the machine, installation setups, and experimental data.
Allowing collaboration among experts and interdisciplinary teams across geographical locations could enhance efficiency, particularly the inspection and analysis, while also establishing a new benchmark for incorporating DTs. 
%It offers real-time synchronization and intuitive interactions, which are beneficial for remote collaboration, particularly 
It is extremely useful when physical travel is restricted, i.e., during the \textit{COVID-19} pandemic. 
%\paragraph{Collaborative Data Exploration and Inspection}
% The main goal is to enhance the AM process by integrating VR and DTs. It enables team members to communicate and collaborate remotely in an immersive and interactive setting. 
Developed features for data exploration and inspection were assessed as useful with real-time synchronization, which is beneficial for team communication and discussion cycles. Scaling and slicing through the data were rated as the most useful features.  
VR provides an immersive way to inspect and analyze complex structures of AM. It is particularly pronounced since it is challenging to perform these tasks on 2D desktop screens. This could lead to better-informed decisions and optimize the design and production accordingly. The spectator mode using conventional input devices was also rated as helpful for some users, e.g., trainees or senior supervisors to promptly join in the discussion.
There were no potential issues regarding VR discomfort or cybersickness observed. They mentioned the framework can be easily adapted by their team although there might be a small learning curve of using this technology.

%\paragraph{Annotation Drawing and Related Features for Team Communication} 
%Regarding features for team collaboration, 
One expert expressed that drawing annotation in VR is intuitive and extremely useful to highlight the point of interest in the data to their collaborators. In this way, they could speed up the process, define multiple regions of defects, and initiate the discussion. 
The experts suggested adding different colors for annotation drawings or each user has a unique color for themselves. 
Besides drawing in the 3D space, they also suggested implementing the annotation drawings on the slicing plane. This would be beneficial to explore the data slice by slice.
Additional features, such as avatar representations and interactions with the virtual whiteboard, were developed to support team communication and discussion.
VR hand representations with animations when pressing the controller buttons were assessed as supportive. The experts suggested adding a help button or tooltips to show the button-mapped functions. However, they stated that it is a learning curve of using the technology.
%%% Additional discussion
The experts highlighted a few challenges of integrating it into their workflow, e.g., logistics and accessibility. They also mentioned about safety while using VR headsets in a limited space in their working setting.
Besides using the framework for inspection and analysis, the experts confirmed employing it for team training and skill development using virtual environments. 

\begin{figure*}[t] %0.323
  \centering
  \includegraphics[width=\linewidth]{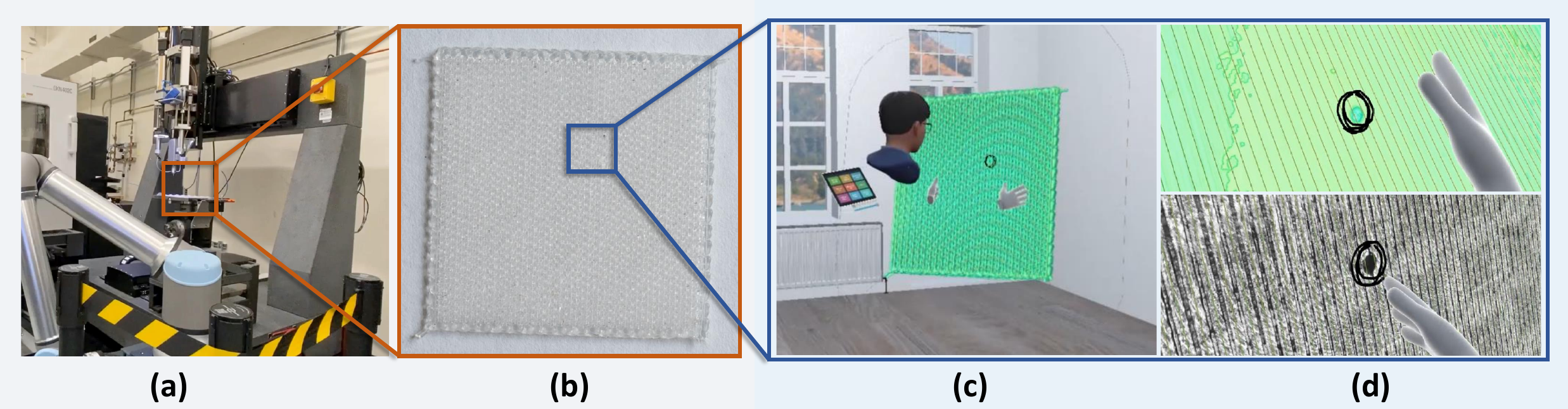}  
  \caption{Inspection and analysis of additively manufactured parts using collaborative VR and DTs: (a) AM and collected multimodal data during the printing process, (b) results of printed part (25\,mm square lattice) in the physical world, (c) users use our proposed VR framework to inspect different multimodal data representations, e.g., marking annotations, and (d) they can compare results between post-curing volume of CT scans (top) and layer-by-layer images during the printing process (bottom).}
  \label{fig:results}
\end{figure*}

\paragraph{Case Study}

In addition to collaborative aspects, utilizing and visualizing multiple data streams of the AM processes was evaluated as essential for diagnostic and enhancing the AM inspection process. 
\autoref{fig:results} shows a case study of using the framework to inspect and analyze AM defects through multimodal data visualization.
A real dataset of additively manufactured parts was used in this study. The size of the physical part was a 25\,mm square lattice.
The two AM experts mentioned that the data can be collected from various sources during the printing process. At the current stage of their workflow, however, there is no best approach to integrate the data together. It is also limited to visualizing and interacting with those data on the desktop screen. 
Hence, it is advantageous to combine and visualize them in an immersive environment. 
It is not only beneficial for exploration but it is crucial for inspecting and defining defects, e.g., evaluating each step and layerwise of the printing process, such as prescribed and machine toolpaths, volume of CT scans, and in-process images.

They stated that it is challenging to define defects via visual inspection or desktop-based systems because the parts can be very small. 
With our VR framework, they can enlarge and inspect them with additional information. 
As an example on this dataset, they found that something poked the first layer of the pre-cure process resulting in defects. 
The framework was helpful by supporting different data modalities, and the data was aligned. They stated that it provides a comprehensive way to explore and inspect the part, e.g., during the pre-cure and post-cure process. Hence, they could compare in-process images with the post-cure volume obtained from CT scans with other modalities overlaid on top, such as prescribed and machine toolpaths. 
In this case, they can see that the toolpaths were actually going through the hole. 
That means the defect was there before the post-cure process, and they could evaluate the layers, which were affected.   
The annotation was helpful in highlighting the defect across multiple modalities.
They could further use the cross-section plane and slicing feature to assess the area.
All interactions were synchronized in real time, and it was beneficial to collaborate and communicate with their collaborators in the same shared environment.
For further improvements, the experts suggested adding additional metadata information to the visualization, such as strut diameters, dispense rates, and scaling units.  
Providing a heatmap for velocity along with the multimodal data visualization could be useful as well. %, and in-process images can be stitched to provide options to explore the whole layer-by-layer image.
Furthermore, they suggested developing a measurement tool in VR to measure the length between two points. 

\section*{Discussion and Future Work}

DTs are increasingly used in manufacturing settings. We tackle the critical needs and explore the use case of emerging DTs, which requires handling multimodal data in an intuitive and collaborative way. 
We demonstrate using our proposed collaborative VR framework for the inspection of AM DTs. 
The proposed framework shows that the traditional desktop-based approach might be not appropriate for team-based collaboration and providing the inspection and analysis intuitively and immersively. Our collaborative VR framework addresses these challenges with advanced visualization and interaction techniques. The framework was evaluated with domain experts in the AM. The results demonstrate the usability, applicability, and potential benefits. All experts found the framework intuitive and effective, and it opens new ways to qualify the AM parts.  

To provide a comprehensive tool for the inspection of AM DTs, we address the challenges ranging from the alignment of multiple data streams collected from the AM processes, multimodal data visualization, occlusion management, and streaming large-scale data to collaborative inspection. 
%%%%%% Alignment
The alignment was rated as helpful to map all data modalities in one place for comparison. It is a crucial component for multimodal data visualization. The alignment can be improved by providing additional information, such as scaling units and a measurement tool to estimate the length of two points. 
%%%%%% Multimodal data visualization
%The experts were impressed with multimodal data visualization. They stated that it was not possible before to visualize those modalities in the immersive environment. %They could intuitively interact with the data and compare between modalities. Visualizing the prescribed and machine toolpaths on top of the volume CT and in-process images was useful to provide insights into behaviors of the defect before and after the post-cure process. They could enlarge and inspect the data layer by layer. 
Different modalities, including sensor data, pressure, velocity, and others, will be incorporated in future work. Investigating visualization techniques for new data streams, such as heatmap and multiscale visualization, would be interesting as well.

%%%%%% Occlusion Management and Volumetric Rendering 
It is essential during the inspection and analysis to consider the occlusion and color maps for volumetric rendering in VR. 
%To provide a global approach to inspecting the inner structures of volumetric data, cutting objects, such as cross-section plane, box, and sphere cutouts, are provided for the users. They can also use the inclusive and exclusive mode to make the inside structures visible or invisible. 
On one hand, it provides an intuitive way to explore volumetric data and grasp the spatial relationships of complex internal structures. On the other hand, the possibility of working as a team to jointly explore data and facilitate collaborative analysis with real-time synchronization is advantageous.
However, there are several challenges in visualizing and rendering volumetric data to avoid discomfort in VR. It includes hardware requirements, which require a powerful computer with a better graphic card to render and maintain a high level of VR immersion. Future research should further investigate occlusion techniques, such as contextual ambient occlusion~\cite{titov2024contextual}, to improve the performance and quality.

%%%%%% Streaming Large-Scale Volumetric Data 
Volumetric data from X-ray CT scans can be extremely large, thus, achieving high-resolution and detailed visualization can be challenging. Moreover, data formats and data transfer for collaborative VR can pose a significant challenge. 
To deal with large-scale and multi-resolution volumetric data, we developed an approach to stream and handle those massive datasets. By utilizing \textit{OpenVisus} for data management, the process of managing and filtering data is enhanced. This could solve the issues of distributing and storing data locally for collaborative users. 
Moreover, dealing with parallel computing is crucial for VR since it is important to maintain the VR rendering while other computations should be processed in the background. Nonetheless, it can be difficult to predict the exact time until it is fully rendered. It also requires significant computational resources to achieve smooth and responsive performance. 
The proposed framework allows users to adjust the quality of the volumetric data. Further investigation with a progressive rendering would be advantageous. Thus, instead of waiting for the entire volume to be rendered, it provides an initial low-resolution and a rough approximation rendering to the user until the computational resources become available.

%%%%%% Collaborative exploration and inspection
DTs in the \textit{Industrial Metaverse} is an emergent concept, which promises various benefits for the DTs ecosystem. It opens new research directions where digital representations and virtual collaboration among users in the immersive environment would become powerful tools for design~\cite{mourtzis2021collaborative}, instruction and maintenance~\cite{thoravi2019loki, oppermann2023industrial}, inspection, optimization, and training~\cite{kuts2020digital, ostrander2020evaluating}. 
It enhances collaboration, interactivity, and team communication via real-time synchronization. Future work can focus on minimizing latency to ensure smooth and synchronous communication between users.
%Ensuring high frame rates for VR and maintaining real-time synchronization for collaborative users is challenging. 
%It requires high-performance computing hardware and good internet connections to ensure smooth and synchronous communication between users.
% The framework ensures the computation for VR logic and rendering is on the main thread to maintain and provide a good user experience in VR. In contrast, other computations run in the background. However, the users might expect longer loading times, and the frame rate might drop while dealing with super high-resolution volumetric data. 
% Multimodal data was integrated into the proposed framework to build part-level DTs. In future work, we aim to explore and develop process-level DTs for full-fidelity with physical-based models and data during the AM printing process.
Since AM DTs involve various properties such as materials, thermal conductivity and resistivity, sensor data, and other physical properties, investigating and leveraging artificial intelligence (AI) and data-driven techniques would be essential for predicting and optimizing data~\cite{tao2018data}. 
For instance, using deep conventional neural networks to extract layer-wise image measurements of the filament thickness and spacing to validate the part-scale reconstruction.
Furthermore, AI-assisted inspection in the virtual environment would be an interesting research direction to assist the users in analyzing, highlighting, and further investigating potential defects.

Apart from the immersive VR environment, team-based collaboration can be applied to situated analytics in a mixed-reality environment of manufacturing settings and across various fields~\cite{tian2023using, lee2021xr}. It could also provide a new approach to incorporate sensor integration of the physical systems to monitor, diagnose, and lead to more accurate and dynamic simulations.
Future work includes increasing data streams, including process-level data of the sensors, and investigating visualization approaches to tackle new data modalities. While the users can communicate in VR, we have to acknowledge that they may have asymmetric knowledge regarding the data. Future work also aims to provide options to save the annotations and results from the inspection as well as their discussion, e.g., drawings on the virtual whiteboard.

\section*{Conclusion}

DTs have become pervasive and increasingly used in industrial manufacturing. While most research focused on automation, 3D modeling, and interoperability of DTs, critical needs of handling and inspecting DTs data with intuitive and immersive approaches remain. 
Moreover, DTs are becoming increasingly complex with various data streams, which often require multiple experts to get involved for effective inspection.
In this work, we explore the use case of DTs in AM part inspection. We present and demonstrate the use of our proposed collaborative VR framework to enable and enhance the inspection process of AM DTs. 
%We present a framework to enhance AM through collaborative VR and multimodal data visualization. 
% While the X-ray CT data is the common standard for AM parts inspection, using desktop-based systems could be time- and resource-intensive and provide limited visualization and interactions compared to the immersive VR environment. 
% Moreover, AM processes and experts are often distributed and scattered across geographical locations. 
% Hence, it is of utmost useful to utilize collaborative VR to support visualization and team collaboration. 
The framework not only focuses on collaborative interactions but also provides several innovative components to improve the inspection and understanding of complex DT data in AM. 
Those components include VR-based interactive alignment, multimodal data visualization, comprehensive occlusion management and rendering, streaming large-scale volumetric data, and features for team-based inspection and collaboration.
Exploratory and semi-structured interviews were conducted with domain experts to evaluate the usability, applicability, and potential benefits. 
A case study of AM inspection using real data collected from the AM process was also presented.
%The feedback was overall positive, and they confirmed the potential benefits and applicability of integrating the framework into their workflow. 
The proposed framework is a promising tool to significantly enhance the inspection process by reducing inspection time and improving efficiency and accuracy. 
It provides a natural way to interact with DT representations by improving spatial awareness, enabling real-time communication, sharing exploration of complex structures, and remote inspection with multiple users.
The framework offers a new benchmark for emerging DTs and pushes the boundaries of current AM DTs inspection methods.
Moreover, it opens new research directions and offers new opportunities for integrating with other DT domains, including the medical, aerospace, and consumer products industries.

\bibliography{biblography}
\section*{Acknowledgements}
This work was performed under the auspices of the U.S.	Department of Energy by Lawrence Livermore National Laboratory under Contract DE-AC52-07NA27344. 
The project has been supported by LLNL LDRD (23-SI-003). 
The work was internally reviewed and released under LLNL-JRNL-861091-DRAFT.

\section*{Author contributions}

V.C., S.N., and G.H. developed the framework. V.C. and H.M. conducted the experiments. R.C., B.A., and B.W. contributed to the user study. V.C. and H.M. prepared the original draft. B.G., P.B., and H.M. provided resources and supervision. All authors reviewed the manuscript. 

\section*{Competing interests}
The authors declare no competing interests.

\section*{Data availability}
Supplementary video is attached. The data generated during and/or analysed during the current study are available from the corresponding author on reasonable request.

\end{document}